\documentclass[manuscript]{acmart}

\usepackage{tabularx}
\usepackage{multirow}
\usepackage{multicol}
\usepackage{balance}

\AtBeginDocument{%
  \providecommand\BibTeX{{%
    \normalfont B\kern-0.5em{\scshape i\kern-0.25em b}\kern-0.8em\TeX}}}

\newif\ifediting
\editingtrue  

\begin{document}

\title[An Ontology of Dark Patterns Knowledge]{An Ontology of Dark Patterns: Foundations, Definitions, and a Structure for Transdisciplinary Action}


\author{Colin M. Gray}
\email{comgray@iu.edu}
\affiliation{
  \institution{Indiana University}
  \city{Bloomington}
  \state{Indiana}
  \country{USA}
}

\author{Nataliia Bielova}
\email{nataliia.bielova@inria.fr}
\affiliation{%
  \institution{Inria Centre at Université Côte d’Azur}
  \country{France}
}
\author{Cristiana Santos}
\email{c.teixeirasantos@uu.nl}
\affiliation{%
  \institution{Utrecht University}
  \country{The Netherlands}
}
\author{Thomas Mildner}
\email{mildner@uni-bremen.de}
\affiliation{%
  \institution{University of Bremen}
  \country{Germany}
}

\renewcommand{\shortauthors}{Gray, Santos, Bielova, \& Mildner}

\begin{abstract}
  Deceptive and coercive design practices are increasingly used by companies to extract profit, harvest data, and limit consumer choice. Dark patterns represent the most common contemporary amalgamation of these problematic practices, connecting designers, technologists, scholars, regulators, and legal professionals in transdisciplinary dialogue. However, a lack of universally accepted definitions across the academic, legislative, and regulatory space has likely limited the impact that scholarship on dark patterns might have in supporting sanctions and evolved design practices. In this paper, we seek to support the development of a shared language of dark patterns, harmonizing ten existing regulatory and academic taxonomies of dark patterns and proposing a three-level ontology with standardized definitions for 64 synthesized dark pattern types across low-, meso-, and high-level patterns. We illustrate how this ontology can support translational research and regulatory action, including transdisciplinary pathways to extend our initial types through new empirical work across application and technology domains.
\end{abstract}

\ccsdesc[500]{Human-centered computing~Human computer interaction (HCI)}
\ccsdesc[500]{Human-centered computing~Empirical studies in HCI}

\keywords{dark patterns, deceptive design, regulation, ontology}

\maketitle

\textcolor{red}{\textbf{Draft: September 15, 2023}}

\section{Introduction}

Deceptive design practices are increasingly common in digital environments, impacting digital experiences on social media~\cite{Schaffner2022-nn,Mildner2021-kp}, e-commerce~\cite{Mathur2019-hx}, mobile devices~\cite{Gunawan2021-pr}, cookie consent banners~\cite{Gray2021-zf}, and gaming~\cite{Zagal2013-gj}, among others. An increasingly dominant framing of these deceptive practices is known as ``dark patterns''\footnote{We use this term to connect our efforts to prior scholarship and legal statute, recognizing that other terms such as ``deceptive design'' or ``manipulative design'' are sometimes used to describe similar tactics. While the ACM Diversity and Inclusion Council has included dark patterns on a \hyperlink{https://www.acm.org/diversity-inclusion/words-matter}{list of potentially problematic terms}, there is no other term currently in use that describes the broad remit of dark patterns practices that include deceptive, manipulative, and coercive patterns that limit user agency and are often hidden to the user.}---describing instances where design choices subvert, impair, or distort the ability of a user to make autonomous and informed choices in relation to digital systems regardless of the designer's intent~\cite{CPRA,DSA2022,FTC2022}. 

While the origins of dark patterns as a concept to describe manipulative design practices goes back over a decade to when the term was coined by practitioner and scholar Harry Brignull~\cite{Brignull_undated-vj}, in the past five years there has been growing momentum in the use of the term to unite scholars, regulators, and designers in transdisciplinary dialogue to identify problematic practices and find ways to prevent or discourage the use of these patterns. According to a recent study of the historical evolution of \#darkpatterns on Twitter by Obi and colleagues~\cite{Obi2022-bd}, since 2019, conversations have included stakeholders not only from design and technology but also social scientists, lawyers, journalists, lawmakers, and members of regulatory bodies and consumer protection organizations. 

Within the regulatory space, in 2022 alone, the term ``dark patterns'' was codified into EU law in the Digital Services Act~\cite{DSA2022}, the Digital Markets Act~\cite{DMA}, and the Data Act proposal~\cite{Data-Act-proposal}, and into US law in the California CPRA~\cite{CPRA}. Regulatory bodies such as the US Federal Trade Commission (FTC), the UK Competition and Market Authority (CMA), the EU Commission, the European Data Protection Board (EDPB) and the the Organisation for Economic Co-operation and Development (OECD) have released guidance on specific types of dark patterns with various levels of overlap with definitions from academic scholarship~\cite{EDPB2022,CMA2022,EUCOM2022,FTC2022, Oecd2022}. In late Summer 2023, the Department of Consumer Affairs in India also released draft guidelines regarding dark patterns~\cite{india_2023-ek}. In addition, the concept of dark patterns has been leveraged in sanctions against companies that have relied upon manipulative practices. Recent actions include a \$245 million USD judgment against Fortnite, a product from Epic Games, for their use of manipulative practices to encourage the purchase of content~\cite{Wodinsky2022-bj} and multiple settlements by various US states against Google for their use of dark patterns to obtain location data~\cite{Press_Release2022-fq,Weatherbed2022-oj}. In the EU, both Data Protection Authorities (DPAs) and court decisions have forbidden certain practices related to dark patterns, including: pre-selection of choices~\cite{CJEU-Planet49-19}; refusing consent if it is more difficult than giving it~\cite{CNILvsGoogle2021,CNILvsFacebook2022}; and misinforming users on the purposes of processing data and how to reject them~\cite{CNILvsFacebook2022, LuxDPA-amazon}. 

As part of this convergent discourse, HCI scholars have addressed the threat of dark patterns in a wide range of publications, proposing definitions and types of dark patterns~\cite{Bosch2016-vc,Gray2018-or,Mathur2019-hx,Mathur2021-rc,Luguri2021-bg}. However, the specific forms that dark patterns can take, the role of context, the ubiquity of the practices, the technologies used or application area, the comparative harms of different patterns, remedies, and the role of user education and countermeasures are still a topic of ongoing research. The consequence of this dynamic topic is of an ever-expanding list of categories and variants whose scale continues to grow.

Two large challenges face an ongoing transdisciplinary engagement with the concept of dark patterns. First, the literature has grown quickly and is siloed, often lacking accurate citation provenance trails of given typologies and definitions, making it difficult to trace where new or more detailed types of patterns emerged and under which conditions. The space that dark patterns scholars have sought to cover is also vast, with important research occurring in specific domains (e.g., games, e-commerce, privacy and data protection) and across different technologies and modalities (e.g., mobile, desktop, conversational user interfaces (CUIs), AR/VR), as shown in a recent systematic review of dark patterns literature~\cite{Gray2023-systematic}. This diversity of research has led some scholars to propose fragmentary, domain-specific typologies without necessarily finding commonalities across domains. Second, regulators---the ultimate decision makers that could provide legal certainty to this landscape--- and policy makers have been interested in the scholarly conversation regarding dark patterns, but have in some cases created wholly new domain-related terminology to describe types already known in the academic literature (e.g.,~\cite{EDPB2023}. In other cases, regulators and policymakers have inconsistently cited academic sources (e.g.,~\cite{FTC2022,EDPB2023}) making connections across the regulatory, legal, and academic spaces fraught. 

We seek to support these challenges and ongoing conversations by building the foundation for a common ontology of dark patterns. By taking the first steps towards building an ontology, we seek to create a shareable, extendable, and reusable knowledge representation of dark patterns. 
This groundwork for an ontology is both domain and application agnostic though it has potential utility in domain or context-specific instances as well. For instance, the {\em Bad Defaults} dark pattern is often embedded in settings menus, pre-set so that users share personal information on social media platforms or accepting to receive advertising content on online shopping sites unknowingly. Such context-specific instances are enabled through {\em Interface Interference}---a domain-agnostic strategy used to manipulate interfaces, privileging certain actions and, thus, limiting discoverability of alternatives. 
As noted by Fonseca~\cite{Fonseca2007-mv}, ontologies can be useful in supporting social science research by ``creating better conceptual schemas and applications.'' To create this preliminary ontology, we build upon ten contemporary taxonomies of dark patterns from both the academic and regulatory literature, and thereafter we identify three levels of hierarchy for pattern types. Hence we harmonize concepts across these taxonomies to provide a consistent and consolidated, shared, and reusable dark patterns ontology for future research, regulatory action, and sanctions. 

We make four contributions in this paper. First, we introduce the hierarchical concepts of low-level, meso-level, and high-level dark patterns to the literature, disambiguating UI-level patterns that may lead to opportunities for detection (low-level) and strategies that may be targeted by policy and legislation (meso- and high-level). Second, by analysing the provenance of dark patterns from academic and regulatory sources, we identify when patterns first emerged and how naming has evolved over time and across sources. Third, we describe a common definition syntax, set of definitions, and hierarchy of dark patterns that aligns disparate terminology from scholars and regulators. Fourth, we demonstrate how the ontology can be strengthened and extended through additional empirical work, and how the ontology can effectively be utilized by practitioners, scholars, regulators, and legal professionals to support transdisciplinary action.

\section{Motivation \& Background}
Since the initial set of a dozen types of dark patterns proposed by Brignull in the 2010s, research has focused on related issues from multiple angles including, but not limited to, e-commerce, games, social media, and IoT~\cite{Gray2023-systematic}. While this scholarship contributes significant insights to the discourse, we noticed varying approaches to adopt existing descriptions, defining novel scenarios in which users are harmed. Meanwhile, the specification of individual typologies creates a certain ambiguity within the overall discourse on the matter. In developing this ontology, we confront numerous timely issues relating to the description of dark patterns, the study of dark patterns and their harms through empirical work, and the leveraging of this scholarship to support legal and regulatory action.

Dark patterns are known to be ubiquitous; however, most pattern types have been explored in relatively narrow contexts or domains with more scholarship needed to fully define causal links, harms, and impacted populations~\cite{Gray2023-systematic}. The HCI community has been engaged and interested in impacting society and the future of technology practices relating to dark patterns~\cite{Gray2023-SIG,Lukoff2021-fk}---and indeed, HCI scholars have been central in the study of dark patterns, revealing insights relating to the harm and severity of dark patterns that then support enforcement action and regulation. However, we currently lack a shared landscape of definitions, types, and language to unify the study of dark patterns. Without this shared landscape, research has become (and will continuously be) fragmented by domain, context, and technology type---which if not addressed, may lead to duplicated effort by scholars working on similar issues in different domains, and additionally may hamper regulatory enforcement due to lack of precision and shared language regarding precisely what dark patterns are used and with what effect. Such lack of a shared ontological framework may also restrict traceability and searchability of dark patterns.

Our work unifies practitioner, scholarly, and regulatory efforts that describe the range of dark patterns, leading to a shared vocabulary and ontology that allows for coordination of efforts across diverse contexts (e.g., technologies, specific functionality, areas of technology use) and stakeholders (e.g., regulators, legal scholars, social scientists, practitioners). This ontology will support not only the advancement of scholarship, but also translational and transdisciplinary efforts that connect scholarship to legal sanctions and regulatory frameworks. For instance, there are now high-level prohibitions of dark patterns by regulatory authorities and legal statute; however, the specific low-level practices that should be deemed illegal under these prohibitions are not yet detailed in enforcement action or case law. This paper connects these different strands of work by harmonizing regulatory and academic work into a single ontology, enabling future scholars from all disciplines to utilise our structures and definitions to support their work.

\section{Methodology}

\begin{figure*}[ht]
  \centering
  \includegraphics[width=\linewidth]{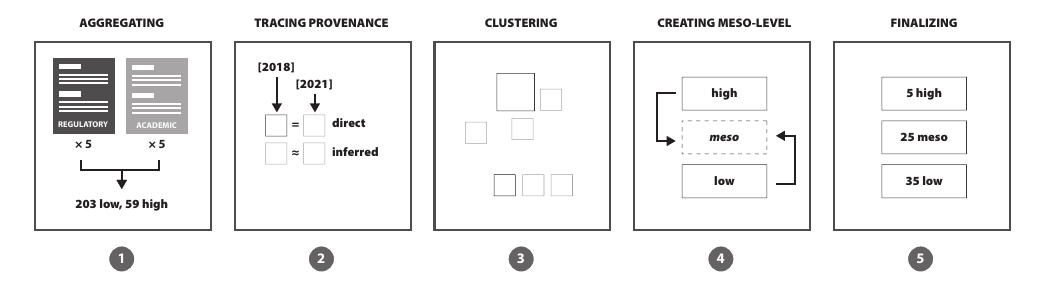}
  \caption{Our method for creating the ontology, mapping to the steps in Section~\ref{sec:ontologymethod}. }
  \Description{Our method for creating the ontology, mapping to the steps in Section~\ref{sec:ontologymethod}. Step 1 shows the consolidation of 10 sources from academic and regulatory literature. Step 2 shows how we traced provenance by first mention and identified inferred and direct mappings. Step 3 shows a visual clustering of pattern names which seemed similar. Step 4 shows our creation of meso-level patterns to map the space between high and low. And Step 5 shows the number of high-level (5), meso-level (25), and low-level (35) patterns.}
  \label{fig:ontologymethod}
\end{figure*}

We used a qualitative content analysis approach~\cite{Hsieh2005-ld} to identify and characterize elements of existing dark patterns taxonomies using the method described in Figure~\ref{fig:ontologymethod}. 

As a research team, we leveraged our collective experiences in human-computer interaction, design, computer science, law, and regulation. Specifically, our team included established dark patterns scholars, including one with a focus on human-computer interaction and design (Authors 1 and 4), one with a focus on computer science and web measurement and experience in regulation (Author 3), and one with a background in computer science and data protection law (Author 2). Across these perspectives, in accordance with previous scholarship, we sought to characterize dark patterns in a transdisciplinary way, drawing on multiple disciplinary perspectives that provide differing views on the origins and types of dark patterns~\cite{Gray2021-zf}. However, these backgrounds also introduce gaps, tensions, and opportunities that relate to the unique experience and academic training of each author. To account for this difference in perspective, each dark pattern type was initially reviewed by each author independently before engaging in conversation amongst the researchers that led to the final agreement on the harmonized type and definition.

\subsection{Data Collection}

We collected dark patterns taxonomies from a total of 10 sources, including: 

\begin{enumerate}
    \item A set of patterns shared on \url{https://darkpatterns.org} since 2010 by Harry Brignull\footnote{This collection of dark patterns was moved to \url{https://www.deceptive.design} in 2022, but the 12 patterns we drew on have been stable since 2018 when the final pattern, ``confirmshaming,'' was added. In 2023, this website was updated to include additional pattern types, resulting in a modified collection of 16 types.}
    \item Scholarly academic sources that present a distinct and comprehensive taxonomy and have either been cited in scholarly and regulatory literature supporting proposed taxonomies or where dark patterns types have been used in regulatory reports without citation but with similar or identical naming~\cite{Bosch2016-vc,Gray2018-or,Luguri2021-bg,Mathur2019-hx}. 
    \item Public reports from stakeholders and regulators in the EU, UK, and USA that include a dark patterns taxonomy~\cite{EDPB2022,EUCOM2022,CMA2022,FTC2022,Oecd2022}. 
\end{enumerate}

The selection of these sources encompass, at the time of our data collection in Fall 2022: i) the most commonly cited taxonomies in the research literature that contributed to regulatory taxonomies in a direct or indirect way, ii) the most comprehensive set of regulatory literature, and iii) all taxonomies cited in regulatory reports, demonstrating an implicit or direct translation from academia to regulation and policy. 

\subsection{Data Analysis}
\label{sec:data-analysis}

Once we gathered the set of taxonomies, we began our analysis by identifying the constitutive components of each taxonomy without considering overlaps across sources through a bottom-up approach. 

\textbf{Quantification of dark pattern types }  
Across the ten taxonomies from academic and regulatory sources collected in Fall 2022, we identified 186 low-level and 59 high-level patterns (a total of 245 patterns). 

After our initial analysis, the patterns used on Brignull's site (\url{https://www.deceptive.design}) were substantially updated in the Summer 2023, and we collected the additional set of patterns for that source---resulting in 11 total sources. Also, the EDPB regulatory report was made final in February 2023, and we used its final taxonomy in this paper after completing our initial mapping in the Fall 2022 based on the draft report taxonomy. Based on the updates to the EDPB guidelines and Brignull's site in the Spring and Summer 2023, the total number of patterns we analyzed included 203 low-level (adding 1 new pattern from the revised EDPB guidelines and 16 patterns from the updated Brignull site) and 59 high-level patterns---\textbf{a total of 262 patterns} (see Tables~\ref{tab:academic} and \ref{tab:regulatory} in the supplemental material).  All taxonomy elements are included in supplemental material for other scholars to build upon.

\textbf{Rationale underlying the high number of dark pattern types} This large number of discrete elements is perhaps unsurprising, since each typology author has used a different point of focus and categorization based on the sector they sought to describe or support. For instance, Mathur et al.~\cite{Mathur2021-rc} and the CMA~\cite{CMA2022} focus on e-commerce; the EDPB focuses on data protection practices within social media platform interfaces~\cite{EDPB2022}, and the FTC~\cite{FTC2022} and EU Commission~\cite{EUCOM2022} focus on guidance specific to their jurisdictions and underlying legal authority. The types themselves also evolved in one case due to input from the  practitioner and regulatory community, which is the case of the EDPB naming of patterns changed slightly from the 2022 draft report to the final 2023 report, with one high-level strategy ``hindering'' changing to ``obstructing'' to bring it into better alignment with academic taxonomies.

\begin{figure*}[ht]
  \centering
  \includegraphics[width=\linewidth]{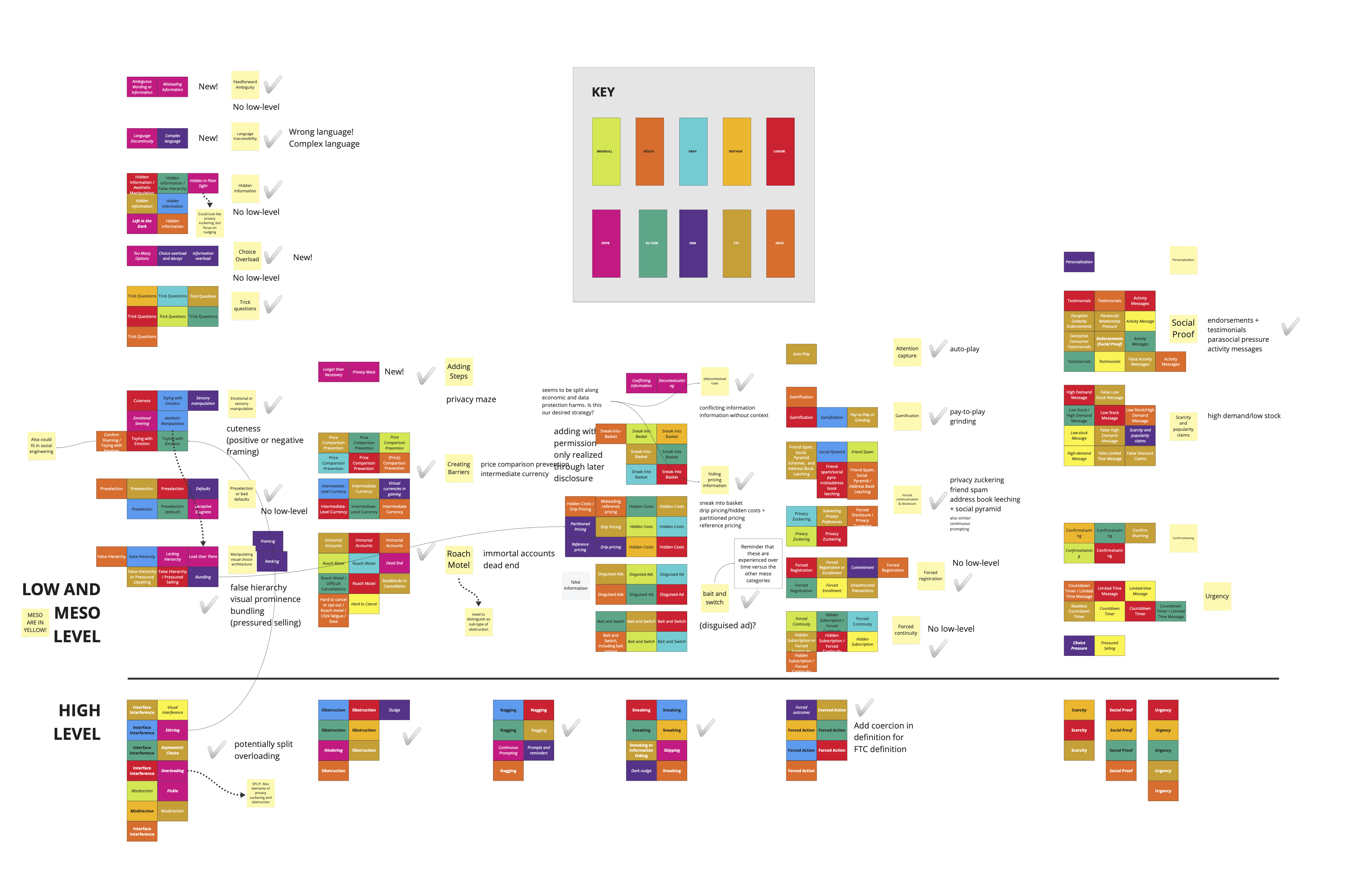}
  \caption{A screenshot of our Miro workspace where we organized and clustered elements of the ten source taxonomies. Columns indicate an entire structure of meso- and low-level patterns underneath a high-level pattern and yellow Post-It notes indicate draft meso-level patterns. The elements are color-coded based on which taxonomy they came from. A full version of this workspace is included as a supplemental material.}
  \Description{A screenshot of our Miro workspace where we organized and clustered elements of the ten source taxonomies. Columns indicate an entire structure of meso- and low-level patterns underneath a high-level pattern and yellow Post-It notes indicate draft meso-level patterns. The elements are color-coded based on which taxonomy they came from.}
  \label{fig:ontologymiro}
\end{figure*}

\subsubsection{Creating the Ontology Framework}
\label{sec:ontologymethod}
We used the following procedure to carefully identify existing taxonomy components, their source, relationships and similarities between components across taxonomies, visualized in Figure~\ref{fig:ontologymethod}:

\begin{enumerate}
    \item \textbf{Aggregating existing patterns.} We first listed all high- and low-level patterns verbatim in the structure originally indicated in the textual source. 
    \textit{High-level patterns} include any instances where the pattern is denoted as a category, strategy, goal, intention, or other parent in a parent-child relationship. \textit{Low-level patterns} indicate specific patterns that are included as a child in a parent-child relationship, or are otherwise undifferentiated in hierarchy (e.g., Brignull's patterns).
    \item \textbf{Identifying provenance through direct citations and inference.} Based on citations provided in the source-document, we indicated any instances where patterns were \textit{directly cited} or otherwise duplicated from previous sources. Because many patterns were uncited---particularly in regulatory reports---we also relied upon citations elsewhere in the document or explicit use of existing pattern vocabulary and definitions from previously published sources, which we indicate as \textit{inferential}. We used these direct and inferential citation patterns to identify where patterns were first introduced, even if they appeared alongside other patterns that had been published previously. This allowed us to map the historical progression of high- and low-level types over time.
    \item \textbf{Clustering similar patterns.} We grouped patterns that appeared either to be identical or similar (in a \textit{is-a} or \textit{equivalent-to} relationship) on Miro (see Figure~\ref{fig:ontologymiro}), using definitions to identify affinities among patterns that did not have identical names. This portion of the analysis was the most extensive, including in depth conversations between an HCI and legal scholars and a careful reading of the definitions as they might be understood by designers and lawyers. We tried out numerous different groupings based on what we understood to be the main focus of each pattern and then sought to characterize what level of pattern each represented.
    \item \textbf{Creating meso-level patterns.} From the findings of this visually-organized analysis procedure, we recognized that there were not only low- and high-level patterns present, but also a \textit{``meso''} level of pattern knowledge. By recognizing similarities among low-level patterns, we introduced \textit{meso-level patterns} into our analysis, identifying these patterns by using the names or elements of existing taxonomies where possible, or coining new names to characterize the low-level patterns we grouped together. If the pattern cluster was specific to low-level UI concerns, we sought to identify a meso-level pattern name that was more abstract and could contain the low-level pattern. If the pattern represented a meso-level abstraction, we did not seek to identify specific low-level instantiations---instead leaving that task for future scholarship efforts in domain- and technology-specific areas.
    \item \textbf{Finalizing the ontology.} Across these three levels of hierarchy, we grouped 233 of the 245 taxonomy elements\footnote{Four ungrouped elements were from the CMA report~\cite{CMA2022} in Fall 2022 and described generic elements of digital systems which were not explicitly framed as deceptive or manipulative: Choice Structure, Choice Information, Feedback, and Messengers. All eight high-level patterns from B\"osch~\cite{Bosch2016-vc} were also excluded since they were not reiterated in any downstream literature.}. After evaluating the changes to the EDPB guideline taxonomy and updated Brignull taxonomy in Spring and Summer 2023, we updated our mapping of 262 patterns, which resulted in no additional novel pattern types. The final ontology includes 5 high-level patterns, 25 meso-level patterns, and 35 low-level patterns---\textbf{a total of 65 patterns}.
\end{enumerate}

\subsubsection{Harmonizing Definitions of Dark Patterns Types}

Building on this ontology framework, we then proceeded to create a definitional syntax across the three levels of the ontology and then created definitions for each final pattern using the following approach: 

\begin{enumerate}
    \item \textbf{Creating definition syntax.} We evaluated the range of approaches to definitions in the existing taxonomies. 
    
    \begin{itemize}
        \item  \textit{Short vs long definitions.} Some definitions were very short (e.g., the EU Commission's definition for \textit{forced registration}: ``Consumer tricked into thinking registration is necessary'') while other definitions were more elaborate (e.g., the FTC's definition for \textit{baseless countdown timer}: ``Creating pressure to buy immediately by showing a fake  countdown clock that just goes away or resets when it times out. Example: `Offer ends in 00:59:48'''; the EU Data Protection Board's definition for \textit{longer than necessary}: ``When users try to activate a control related to data protection, the user journey is made in a  way that requires more steps from users, than the number of steps necessary for the activation of data invasive options. This is likely to discourage them from activating such control.''). 
        
        \item  \textit{Description of the definitions.} Most definitions were based in a description of user interaction with a system, like the examples above; however, Brignull's 2018 definitions were written in first-person language demonstrating how a user would experience a dark pattern (e.g., the definition for \textit{roach motel}: ``You get into a situation very easily, but then you find it is hard to get out of it (e.g. a premium subscription).'') Interestingly, Brignull's 2023 language appears to model other taxonomies with all definitions beginning with ``The user struggles...,'' ``The user expects...'' or similar structures. 
    
       \item \textit{Definition structure and syntax.} We used an iterative process where two authors independently and collaboratively tested different definition structures. Based on these efforts and through discussion, we finalized sample definition structures and syntax that captured the relevant type of knowledge (e.g., strategy, angle of attack, means of execution). For instance, all high-level patterns included the interplay of an undesired action and a limitation of their decision-making or free choice. Meso-level patterns addressed a mismatch in users' expectations of a system and the relevant impact. Low-level patterns identified how they manifest their parent high- and meso-level pattern in relation to one or more elements of the UI and a mismatch of expectation and resulting effect on the user experience.
    \end{itemize}

    \item \textbf{Creating and evaluating high- and meso-level pattern definitions.} We then drafted definitions for all high- and meso-level patterns, iterating on the structure until we found a syntax that appeared to address all critical elements of the existing definitions and allow us to clearly indicate how the pattern subverted user autonomy and manifest as deceptive or coercive. We began with definitions at these levels since low-level patterns were already grounded in specific UI examples, and thus more effort was needed to identify what components a definition at a higher level of abstraction should include. Regarding evaluation, our set of 30 definitions and the draft definition structures were then shared via an open Google Doc with members of a large Slack community focused on research and enforcement action relating to dark patterns. We asked this community for feedback on the utility of the definitions, the completeness of the definition structures, and the ability of these definitions to leave as open-ended the many different low-level manifestations of dark patterns. Multiple community members gave us feedback which allowed us to validate the general face validity of the definitions, and we continued to iterate on our structure and language in response.
    \item \textbf{Finalizing low-level pattern definitions.} After mapping out the initial 30 definitions, we created definitions for the 34 low-level patterns that were grounded in the specifics of the UI execution. These patterns were easier to write since many taxonomy definitions (in particular those from Brignull~\cite{Brignull_2023}, Gray~\cite{Gray2018-or}, and the FTC~\cite{FTC2022}) included richer detail for patterns that pointed towards a real-world implementation. As a research team, we read and edited the definitions until we were satisfied with their level of consistency and relationships to the higher-level categories in which they belonged. All definitions are included in the appendix of this paper and supplemental materials to support future work.
\end{enumerate}

\section{Mapping the Evolution of Dark Patterns}
\label{sec:evolution}

\begin{figure*}[ht]
  \centering
  \includegraphics[width=\linewidth]{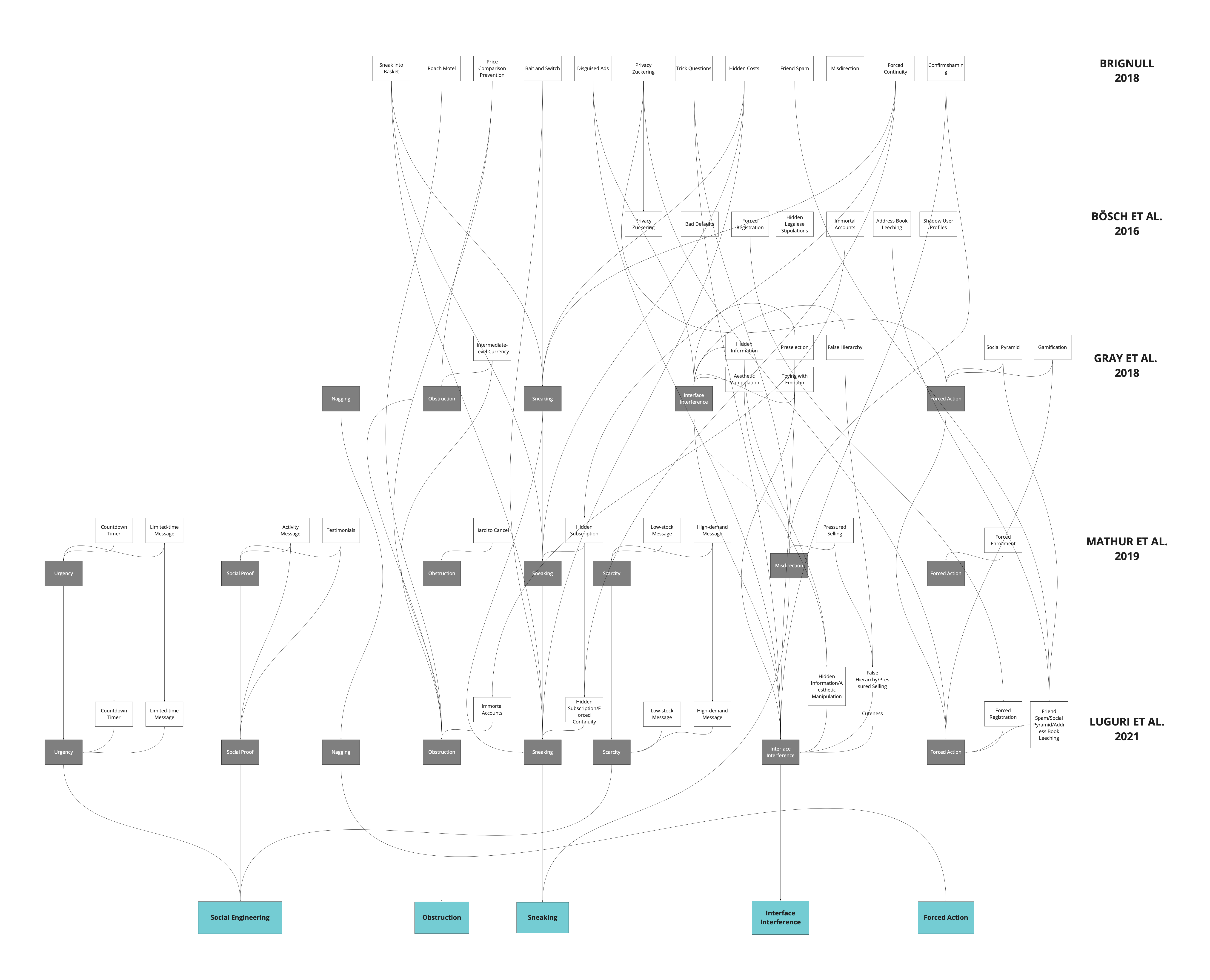}
  \caption{A visual mapping of the evolution of dark patterns in the academic taxonomies we analyzed from 2018-2021. Each row includes elements of the related taxonomy by year and source, and connecting lines indicate relationships between or reiterations of different patterns over time. Pattern names in gray boxes are high-level patterns, pattern names in white boxes are low-level patterns or otherwise lack hierarchy, and pattern names at the bottom are the final high-level patterns we adopt in our ontology. A full version of this mapping is included as a supplemental material.}
  \Description{A visual mapping of the evolution of dark patterns in the academic taxonomies we analyzed from 2018-2021. Each row includes elements of the related taxonomy by year and source, and connecting lines indicate relationships between or reiterations of different patterns over time. Pattern names in gray boxes are high-level patterns while pattern names in white boxes are low-level patterns or otherwise lack hierarchy. A full version of this mapping is included as a supplemental material.}
  \label{fig:ontologyevolution}
\end{figure*}

Pattern names have largely stabilized in the past five years, including high-level pattern types (e.g., nagging, obstruction, sneaking, interface interference, forced action) and low-level patterns (including Brignull's~\cite{Brignull_undated-vj,Brignull_2023} and those introduced by Gray et al.~\cite{Gray2018-or} and Mathur et al.~\cite{Mathur2019-hx}). A mapping of these patterns over time across the academic and practitioner sources we considered is included in Figure~\ref{fig:ontologyevolution}.

\textbf{High-level patterns} were most likely to co-occur across multiple sources. For instance, Gray et al.'s~\cite{Gray2018-or} original five high-level ``dark pattern strategies'' were found across multiple other sources, even if they were not consistently cited: nagging \cite{Luguri2021-bg,EUCOM2022}, obstruction \cite{Mathur2019-hx,Luguri2021-bg,EUCOM2022,FTC2022}, sneaking \cite{Mathur2019-hx,Luguri2021-bg,EUCOM2022,FTC2022}, interface interference \cite{Luguri2021-bg,EUCOM2022,FTC2022}, and forced action \cite{Mathur2019-hx,Luguri2021-bg,EUCOM2022,FTC2022} (FTC uses ``coerced action'' instead). As shown in Figure~\ref{fig:ontologyevolution}, virtually all of the high level patterns proposed by Gray et al. in 2018 were carried forward in other academic taxonomies. 
In Brignull's 2023 changes to \url{https://www.deceptive.design}, multiple high-level strategies from Gray et al.'s~\cite{Gray2018-or} taxonomy were added to the website (nagging, obstruction, sneaking, forced action, visual interference)---however, these changes were not cited and Brignull continued his practice of not providing direct citations or hierarchical structure to his patterns. After their introduction in Mathur et al.~\cite{Mathur2019-hx}, newly introduced categories relating to social psychology or behavioral economics also became common: urgency \cite{Luguri2021-bg,EUCOM2022,FTC2022}, scarcity \cite{Luguri2021-bg,FTC2022}, and social proof \cite{Luguri2021-bg,EUCOM2022,FTC2022} (the FTC bundles ``Endorsements'' with ``social proof''). We have grouped these types together as part of a sixth high-level pattern of ``social engineering.''

\textbf{Domain or context-specific patterns.} The most volatility has occurred in relation to \textit{domain-} or \textit{context-specific patterns}. These include expansions of Mathur et al.'s~\cite{Mathur2019-hx} high-level patterns of ``social proof'' and ``scarcity,'' which have since been reiterated by the EU Commission~\cite{EUCOM2022} and OECD~\cite{Oecd2022} and extended by the CMA~\cite{CMA2022} and FTC~\cite{FTC2022} taxonomies. In addition, the EDPB guidance on dark patterns in social media~\cite{EDPB2022} included a wholly new set of 6 high-level and 15 low-level patterns, although the majority of these could be inferred as similar to already existing patterns proposed in the academic literature. Importantly, though, the EDPB taxonomy included multiple patterns which we found to be new low-level or meso-level additions, including ``privacy maze,'' ``dead end,'' ``conflicting information,'' ``information without context'' (which we renamed from the EDPB pattern ``decontextualizing''), and ``visual prominence'' (which we renamed from the EDPB pattern ``look over there''). Similarly, the CMA taxonomy focused on choice architecture as a guiding structure with three categories focused on choice ``structure,'' ``information,'' and ``pressure.'' This taxonomy structure also yielded new patterns, including ``bundling,'' ``complex language,'' and ``personalization.''

Our analysis demonstrates the value in classifying or generating context-specific patterns that illuminate gaps in current taxonomies, and also the benefit of mapping these patterns within larger ontologies to identify abstractions of patterns that may apply across many domains, contexts, and legal fields. Our final ontology mapping is included in Figures~\ref{fig:ontologypt1} and \ref{fig:ontologypt2} and can also be found in the supplementary materials.

\begin{figure*}[ht]
  \centering
  \includegraphics[width=\linewidth]{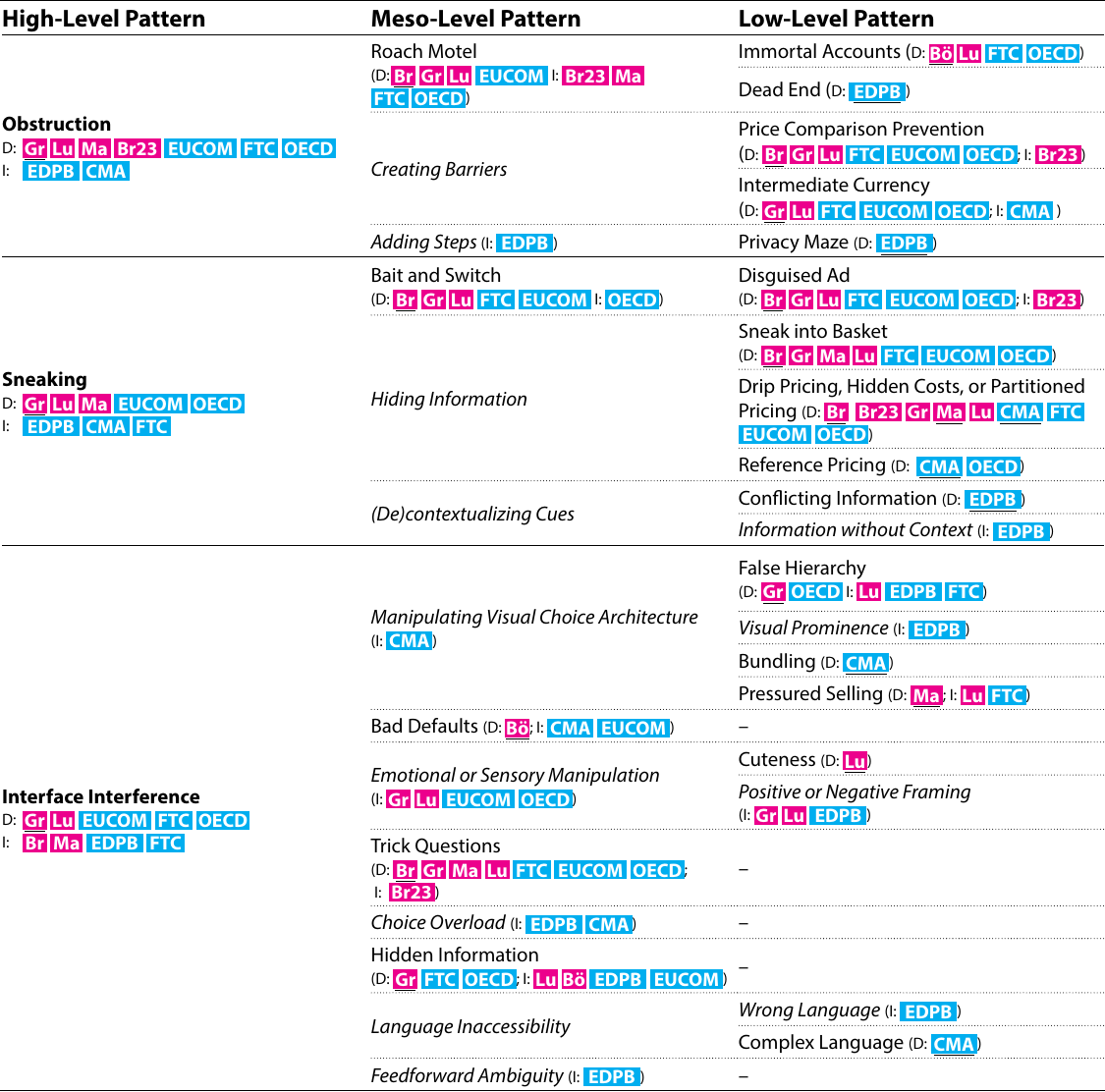}
  \caption{Our ontology of dark patterns organized by level of pattern. ``D'' indicates a \textit{direct} use of the pattern language in the original source(s) and ``I'' indicates an \textit{inferred} similarity between different terminology used across two or more pattern types. Sources are indicated by abbreviation and are colored \textcolor{cyan}{cyan} if they are regulatory reports or \textcolor{magenta}{magenta} if they are academic or practitioner sources. ``Br'' indicates his 2018 patterns and ``Br23'' indicates his 2023 patterns. \textit{Italized pattern names} indicate new pattern types introduced in this paper while all other text relies upon the sources indicated. \underline{Underlined sources} indicate the earliest mention of that pattern or patterns in the sources we analyzed. A full description of the inferred pattern names is included in supplemental material to support future work.}
  \Description{This table organizes patterns in columns that include high-level, meso-level, and low-level pattern types. Each row represents a broad high-level pattern and matching meso-level and low-level patterns. Each pattern name includes indications of which regulatory and/or academic literature maps to that pattern. A full report of these mappings is included in a machine-readable xslx file as part of the supplemental documents.}
  \label{fig:ontologypt1}
\end{figure*}

\begin{figure*}[ht]
  \centering
  \includegraphics[width=\linewidth]{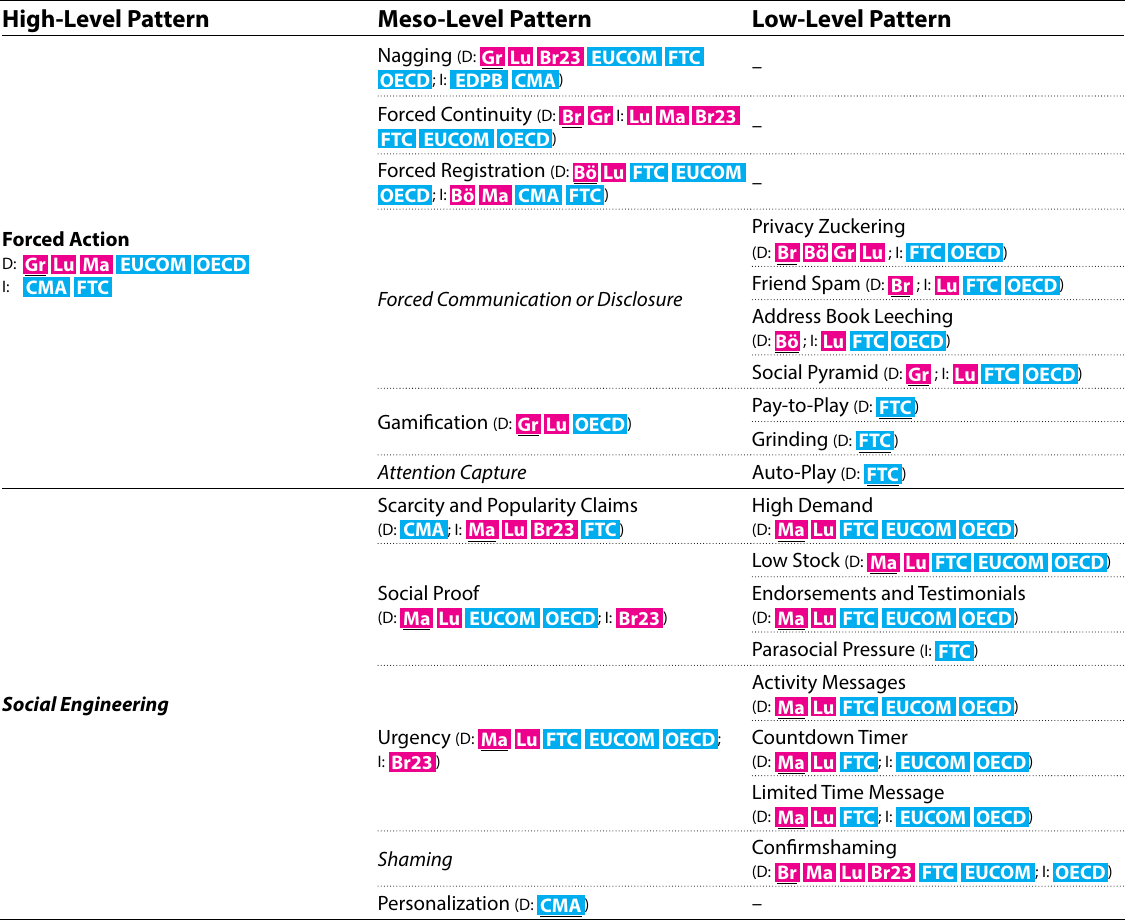}
  \caption{Ontology of dark patterns organized by level of pattern, continued.}
  \label{fig:ontologypt2}
\end{figure*}

\section{Creating a Definitional Structure by Ontology Level}

As described in Section~\ref{sec:ontologymethod}, our ontology includes three different levels of hierarchy: 

\begin{itemize}
    \item \textbf{High-level patterns} are the most abstracted form of knowledge, including general \textit{strategies} that characterize the inclusion of manipulative, coercive, or deceptive elements that might limit user autonomy and  decision making. These patterns are context-agnostic and can be employed through a range of modalities and technologies (e.g., desktop, mobile, VUIs, VR/AR) and application types (e.g., e-commerce, gaming, social media). 
    \item \textbf{Meso-level patterns} bridge high- and low-level forms of knowledge and describe an \textit{angle of attack} or specific approach to limiting, impairing, or undermining the ability of the user to make autonomous and informed decisions or choices. These patterns are content-agnostic and may be interpreted in a contextually-appropriate way based on the specific context of use or application type. 
    \item \textbf{Low-level patterns} are the most situated and contextually dependent form of knowledge, including specific \textit{means of execution} that limits or undermines user autonomy and decision making, is described in visual and/or temporal form(s), and is likely to be detectable through algorithmic, manual, or other technical means.
\end{itemize}

To create a definitional structure for each level, we first used a subset of approximately ten dark patterns types and definitions in order to ``play-test'' a combined and unified definition for dark patterns types at multiple levels of granularity (i.e., high, meso, low). Through this process, we considered not only the level of abstraction inherent in dark patterns at differing levels, but also the interaction between: the user's expectations of what should or would be likely to occur (i.e., manipulation of the gulf of execution); the user's identification that something had occurred that they did not wish to happen (i.e., manipulation of the gulf of evaluation); and the mechanisms used to inform or execute manipulation in either of these prior elements. We also considered cases where the deception or manipulation was likely to be hidden to the user (e.g., cases of sneaking, obstruction, or interface interference) as well as cases where deception or coercion was overt and known to the user (e.g., forced action). Based on this iterative generation of a definitional structure, we created a standardized syntax for each dark pattern level, described below. All 65 final definitions are included as a supplemental material.

\subsection{High-Level Patterns}

\begin{quote}
    \textbf{\{HIGH-LEVEL DARK PATTERN\}} is a strategy which \textbf{\{UNDESIRED ACTION\}} that [optionally, if known to users, would] \textbf{\{DISTORT/SUBVERT/IMPEDE/OTHERWISE LIMIT USERS’ AUTONOMY, DECISION-MAKING, OR FREE CHOICE\}}.
\end{quote}

Across our 5 high-level pattern definitions, we considered {\em undesired actions} such as: hiding, disguising, delaying, redirecting, repeating, impeding, privileging, or requiring actions. We also considered a range of \textit{mechanisms} that could be used to limit users' autonomy, decision-making, or free choice such as: foregrounding unrelated tasks, dissuading a user from taking an action, confusing the user, limiting discoverability of action possibilities, causing a user to unintentionally take an action they would likely object to, or forcing a user to take an action they would not otherwise take. Most of these definitions placed a focus on mechanisms which were primarily hidden, resulting in the user being deceived, such as: ``\textit{Interface Interference} is a strategy which privileges specific actions over others through manipulation of the user interface, thereby confusing the user or limiting discoverability of relevant action possibilities.'' However, the definition for \textit{Forced Action} was focused more on the coercive nature of the interaction which may involve users' awareness they are being manipulated: ``\textit{Forced Action} is a strategy which requires users to perform an additional and/or tangential action or information to access (or continue to access) specific functionality, preventing them from continuing their interaction with a system without performing that action.''

\subsection{Meso-Level Patterns}

\begin{quote}
    \textbf{\{MESO-LEVEL DARK PATTERN\}} subverts the user's expectation that \textbf{\{EXPECTATION\}}, instead producing or informing \textbf{\{DIFFERENT EFFECT ON USER\}}.
\end{quote}

Across our 24 meso-level pattern definitions, we considered a range of \textit{user expectations} such as: presence of relevant and timely information, match between user goal and action, completeness and truthfulness of information provided, and the ability to change one's mind and reverse a decision. We also considered a range of \textit{potential negative effects} on the user, such as: unexpected or unanticipated outcomes, confusion or pressure, being prevented from locating relevant information, or making a different choice than they would otherwise make. Meso-level definitions as a set touched on many different aspects of the user experience, with some pointing more towards static moments in the user journey and others describing temporal effects that might be realized over a longer portion of the user journey. For instance, these two patterns represent instances where the focus was primarily on static UI elements or a particular moment of interaction:

\begin{itemize}
    \item ``\textit{Manipulating Visual Choice Architecture} subverts the user’s expectation that the options presented will support their desired goal, instead including an order or structure of options that makes another outcome more likely.''
    \item ``\textit{Scarcity or Popularity Claims} subverts the user’s expectation that information provided about a product’s availability or desirability is accurate , instead pressuring the user to purchase a product without additional reflection or verification.''
\end{itemize}

In contrast, other patterns represented instances where the full effect of the pattern was felt over time and might involve multiple interactions with a system that accumulate to achieve the overall effect:

\begin{itemize}
    \item ``\textit{Roach Motel} subverts the user’s expectation that an action will be as easy to reverse as it is to make, instead creating a situation that is easy to get into, but difficult to get out of.''
    \item ``\textit{Hiding Information} subverts the user’s expectation that all relevant information to make an informed choice will be available to them, instead hiding information or delaying the disclosure of information until later in the user journey that may have led to them making another choice.''
\end{itemize}

\subsection{Low-Level Patterns}

\begin{quote}
    \textbf{\{LOW-LEVEL DARK PATTERN\}} uses \textbf{\{RELATED HIGH- AND MESO-LEVEL DARK PATTERN\}} to \textbf{\{ELEMENT OF UI ALTERED\}}. As a result, \textbf{\{INCORRECT USER EXPECTATION\}} leads to \textbf{\{UNDESIRED EFFECT ON USER\}}.
\end{quote}

Across our 35 low-level definitions, we considered a range of \textit{means of execution} in the UI or user experience, such as: provision of information that is conflicting, prohibiting certain kinds of interactions, adding items without a user's knowledge, providing incomplete or misleading information, distracting a user through extraneous cues, or using social or other extrinsic pressure to steer user's decisions. These means of execution were supported by a wide range of \textit{incorrect user expectations and related undesired effects}, including: preventing a user from making an informed choice about their privacy or purchase of a product, disclosing incomplete or misleading information that leads to choices the user would not otherwise make, or distracting a user and thus preventing them from discovering information that would be relevant to their decision. Low-level patterns all exploit the user experience in direct ways, but address different aspects of the experience:

\begin{itemize}
    \item Focus on specific user interactions that are limited (e.g., ``\textit{Price Comparison Prevention} Creates Barriers and uses Obstruction by excluding relevant information, limiting the ability of a user to copy/paste, or otherwise inhibiting a user from comparing prices across two or more vendors. As a result, the user cannot make an informed decision about where to buy a product or service.'')
    \item Focus on a coordinated set of user interactions that produce the desired effect (e.g., ``\textit{Privacy Mazes} Add Steps and use Obstruction to require a user to navigate through many pages a result, the user is prevented from easily discovering relevant information or action possibilities, leaving them unable to make informed decisions regarding their privacy.'')
    \item Focus on discrete UI elements (e.g., ``\textit{False Hierarchy} Manipulates the Visual Choice Architecture, using Interface Interference to give one or more options visual or interactive prominence over others, particularly where items should be in parallel rather than hierarchical. As a result, the user may misunderstand or be unable to accurately compare their options, making a selection based on a false or incomplete choice architecture.'')
    \item Focus on user comprehension of the interface (e.g., ``\textit{Wrong Language} leverages Language Accessibility, using Interface Interference to provide important information in a different language than the official language of the country where users live. As a result, the user will not have access to relevant information about their interaction with the system and their ability to choose, leading to uninformed decisions.'')
\end{itemize}

\section{Extending the Ontology Based on Current and Future Scholarship}

Dark patterns researchers have addressed the impact of manipulative, deceptive, and coercive design in a range of technological domains. While these efforts are important in protecting online users and identifying areas for regulatory or legal impact, the novelty and breadth of this work potentially hinders an exhaustive mapping of dark patterns onto our ontology. Building on our proposed ontology, we identify pathways for many stakeholders to contribute to the growth of ontology elements---both through the addition of new patterns and strengthening contextual or domain-specific examples of existing patterns. This extension can help not only to anchor instances of patterns from future studies in existing literature, but also to enable the scholarly community to extend or further characterize these pattern types. The ontology's stratification allows anyone to extend the current framework by following the structure and syntax given for each high, meso, and low level dark pattern type. 

To perform this mapping and extension exercise, we sought to identify existing alignment between proposed dark patterns and the ontology and also consider how a source might offer new perspectives or novel examples of dark patterns. The method we used to extend the ontology involves three steps:

\begin{enumerate}
    \item We analyzed the dark pattern definition included by the author and, if provided, considered any cited relationships to other dark patterns and related terminologies.
    \item We then aligned the author's definition with the syntax of the high, meso, and low levels, placing the dark pattern at the most logical level of abstraction.
    \item Finally, we considered how the addition of the type informs a revision of the ontology. A type could reiterate an existing type in the ontology (leaving the core ontology unchanged), extend an existing type in the ontology (providing rationale for a more expansive definition of an existing type), or identify the presence of a wholly new type (adding a type to the core ontology).
\end{enumerate}

 This section demonstrates how we envision for the community to extend the ontology by drawing examples from three contemporary studies defining dark patterns from domain and context-specific areas, underlining the decision behind selecting these relevant work. We also show how the ontology can be extended to map legislation and case law relating to dark patterns. Table~\ref{tab:extending-ontology} summarizes how three different sources were compared to our ontology through this method, demonstrating how the ontology could be extended by the community over time. We plan to host the ontology on a website which will include community-vetted changes over time that follow this process in a public, deliberative manner. Future versions of the ontology will also be versioned and include a version history for citation accuracy.

\begin{table*}[ht!]
\centering
\renewcommand{\arraystretch}{1.4}
\setlength{\tabcolsep}{5pt}
\begin{tabular}{p{.22\textwidth}p{.38\textwidth}p{.2\textwidth}p{.1\textwidth}}
\toprule
\multicolumn{4}{p{0.95\textwidth}}{\centering\textbf{Extending the Ontology}} \\
\toprule
\textbf{Name} & \textbf{Definition from the Sources} & \textbf{Mapping to Ontology} & \textbf{Level}\\ \midrule

\textit{Linguistic Dead-End}~\cite{hidaka2023-la} & ``[D]esign patterns wherein language use prevents or makes it very difficult for the user to understand crucial functionality [...]''. & \textit{Language Inaccessibility} & extends meso-level\\ 
\textit{Untranslation}~\cite{hidaka2023-la} & ``[D]esign patterns in which part or all of the app is in a language unfamiliar to the people using it, even if the app is stated as available in the local language in the store''. & \textit{Wrong Language} & extends low-level\\ 
\textit{Alphabet Soup}~\cite{hidaka2023-la} & ``[D]esign pattern  language use prevents or makes it very difficult for the user to understand crucial functionality [...]''. &  \textit{Language Inaccessibility} & new low-level\\ \midrule

\textit{Extraneous Badges}~\cite{Gunawan2021-pr} & ``[D]esign elements --- often tiny, brightly colored circles---that visually highlight UI elements that require immediate user attention''. &  \textit{Aesthetic Manipulation} & new low-level\\
\textit{Account Deletion Roadblocks}~\cite{Gunawan2021-pr} 
& ``\textit{Unclear deactivation/deletion options} covers cases where a service insufficiently communicates what will happen if a person deactivates or deletes their account.'' & \textit{Roach Motel} & new low-level\\ 
& ``\textit{Time-Delayed Account Deletion} covers cases where a service will only initiate the account deletion process after a cool-off period, rather than instantaneously.'' & \textit{Roach Motel} & new low-level\\ \midrule

\textit{Engaging Strategies}~\cite{Mildner2023-ab} & ``[D]ark patterns where the goal is to keep users occupied and entertained for as long as possible''. & \textit{Social Engineering} & extends high-level\\ 
\textit{Governing Strategies}~\cite{Mildner2023-ab} & Dark patterns ``that navigate users' decision-making towards the designers' and/or platform providers' goals''. & \textit{Obstruction} & extends high-level\\ 
\textit{Labyrinthine Navigation}~\cite{Mildner2023-ab} & ``[N]ested interfaces that are easy to get lost in, disabling users from choosing preferred settings''. & \textit{Privacy Maze} & extends low-level\\
\bottomrule
\end{tabular}
\caption{This table presents an overview of selected dark patterns from Hidaka et al.~\cite{hidaka2023-la}, Gunawan et al.~\cite{Gunawan2021-pr}, and Mildner et al.~\cite{Mildner2023-ab} to demonstrate extending the dark pattern ontology.}
\label{tab:extending-ontology}
\end{table*}

\subsection{Dark Patterns In Non-Western Cultural Contexts}
Hidaka et al.~\cite{hidaka2023-la} studied dark patterns in Japanese apps and identified two dark pattern types---``untranslation'' and ``alphabet soup''---which are sub-types of a novel ``linguistic dead-end'' dark pattern. We closely evaluated the authors' definition of \textit{Linguistic Dead-End}, where use of a foreign language hinder users from understanding the consequences of their interactions. When comparing these three patterns to our ontology, the high-level pattern \textit{Linguistic Dead-End} appears to fit within the existing meso-level dark pattern \textit{Language Inaccessibility}, while extending its coverage. The remaining two low-level patterns, \textit{Untranslation} and \textit{Alphabet Soup}, can then be nested as two low-level types underneath the same meso-level dark pattern, with \textit{Untranslation} mapping to and extending the existing \textit{Wrong Language} dark pattern and \textit{Alphabet Soup} forming a new low-level pattern. In this case, the three dark patterns extend and further support a distinct area of the ontology, demonstrating how novel contexts help to usefully supplement existing dark patterns and identify new low-level means of execution.

\subsection{Contextual Dark Patterns in Different Screen Modalities}
Gunawan et al~\cite{Gunawan2021-pr} investigated the presence of dark patterns across different screen modalities, describing eight novel dark pattern types which limit the choices of users depending on the device used. In the provided definitions for each proposed dark pattern, the authors included links to previously defined dark patterns---linking these patterns to elements of the ontology, thus providing an easy mapping path. The \textit{Extraneous Badges} dark pattern, for example, is indicated as related to \textit{Aesthetic Manipulation}~\cite{Gray2018-or} as a form of \textit{Interface Interference}, and would result in this dark pattern being included as a new low-level type in the ontology. Similarly, using the authors' definitions and identification of mapping in the paper text, \textit{Account Deletion Roadblocks} could extend \textit{Roach Motel} through two specific new low-level types focusing variously on insufficient communication and time delay: \textit{Unclear Deactivation/Deletion Options} and \textit{Time-Delayed Account Deletion}. These examples illustrate how contextual and situational links to previously defined dark patterns support the ontology, describing specific situations that strengthen established dark patterns and identify new low-level means of execution.

\subsection{Domain-Specific Dark Patterns in Social Media Applications}
Mildner et al.~\cite{Mildner2023-ab} investigated dark patterns on social media platforms, proposing five dark patterns across two strategies. As with Hidaka et al., the granularity of their definitions imply a mapping on multiple levels of the ontology. We began by drawing from the authors' definitions of \textit{Engaging Strategies} and \textit{Governing Strategies}. The authors describe the aim of \textit{Engaging Strategies} as entertaining users for as long as possible, related to \textit{Attention Capture}~\cite{monge_roffarello2023-ac}, which is already included in the ontology as a meso-level pattern under \textit{Forced Action}. However, some elements of the original definition (e.g., occupying and entertaining) fit more closely within concepts of \textit{Social Engineering}. Similarly, \textit{Governing Strategies} can be partially linked to multiple patterns in the ontology. For example, as the authors originally suggest, the strategy can be enabled through \textit{Interface Interference}. However, \textit{Governing Strategies} also offers a high-level focus to inspect \textit{Obstruction} with \textit{Labyrinthine Navigation}, presenting an interesting adaption of \textit{Privacy Maze} already present in the ontology. These examples indicate how the authors make their dark pattern types distinct from prior ones, functioning as a lens that might invite reinspection of dark patterns in the ontology and perhaps indicate opportunities for further development of low-level patterns. 

\subsection{Dark Patterns in Legislation and Case Law}
An alignment between legislation, the ontology, and case law shows that it could also be a robust and reliable artifact for regulators and policy makers to use in their compliance monitoring and enforcement actions. 

\textbf{Mapping the ontology to case law}
Dark patterns have been detected in regulatory cases by enforcers, such as Data Protection Authorities (DPAs) and Consumer Protection Authorities, for more than a decade~\cite{FTC2022,EUCOM2022}. However few cases explicitly designate dark patterns as such.\footnote{Case law and legal frameworks have recently been added to the \url{https://deceptive.design} site, which include mappings to specific dark patterns~\cite{deceptivedesign-legalcases}.} Decisions analyse several practices that are related to dark patterns, but without qualifying each practice into a concrete granular type of dark pattern. Current case law descriptions of the use of dark patterns often report infringements only at a general level, but without qualifying each practice as a concrete type of dark pattern~\cite{Santos_Rossi_2023}. In doing so, case law could miss lower-level granularity that may translate across domains. A recent example shows that a EU regulator, the Italian DPA, used the concept of dark patterns related to certain consent practices for the first time in an official EU legal decision~\cite{ItalianDPAvsEdiscom2023}. By mapping case law to the ontology, regulators can gain additional knowledge identifying where dark patterns practices at multiple levels and in multiple combinations are at play, and were deemed to be illegal per jurisdiction~\cite{Leiser-2022-regpluralismGDPR/UCPD}, enhancing legal certainty about dark patterns practices. For example, the EU Court of Justice has ruled that the practice called ``pre-selection'' violates the GDPR~\cite{CJEU-Planet49-19}, which maps to the meso-level dark pattern \textit{``Bad Defaults''} in our ontology. 

Further, the ontology has the potential to support enforcement decisions since it can test and confirm the traceability of concrete dark patterns-related practices. For instance, the Italian Data Protection Authority has already added the keyword “dark pattern” to the available tags of their online database\footnote{\url{https://www.garanteprivacy.it/temi/internet-e-nuove-tecnologie/dark-pattern}}---a useful effort that should be extended to official and unofficial searchable databases of enforcers’ decisions. Connecting case law to multiple levels of dark patterns in our proposed ontology has the potential to inform enforcers of different jurisdictions in the EU/US and reduce the risks of gaps or overlaps. 

\textbf{Mapping the ontology to legislation}
The proposed ontology can also help regulators across different jurisdictions to understand relationships between different definitions of dark patterns, including high-, meso- and low level dark patterns, including when such definitions map to existing and upcoming legislation. The recent EU Digital Service Act (DSA)\cite[Art.25(3)(b), recital 67]{DSA2022} explicitly prohibits user manipulation and specifies that further guidelines will be given on a specific practice, 
where ``repeatedly requesting that the recipient of the service make a choice where that choice has already been made, especially by presenting pop-ups that interfere with the user experience''; this example maps well to the proposed {\em Nagging} dark pattern in our ontology. Because new legislation, such as the DSA\cite{DSA2022}, Data Market Act (DMA)\cite{DMA}, Data Act \cite{Data-Act-proposal}, and California CPRA~\cite{CPRA} contain dark patterns specific prohibitions, we believe the proposed ontology has the capability to ensure a precise mapping between the concepts of dark patterns in  research literature and the legally-binding provisions. When the concepts of the ontology are mapped to a legal concept, then it is easier for regulators to link a specific dark pattern to a concrete binding legislative provision. Consequently, the ontology will help to conclude the normative value of such practice---whether a specific dark pattern is illegal or legal---and what relevant obligations and rights are derived from the law and must be enforced. If regulators and policy-makers across jurisdictions rely on the same definitions of dark patterns, this can assure an easier re-use of case law for future legal cases. 

\section{Using the Ontology to Support Transdisciplinary Engagement}

In this ontology, we seek to synthesize and harmonize existing academic and regulatory taxonomies while adding useful and consistent structure to allow for other stakeholders to build upon and derive benefit from a shared description of dark patterns knowledge. This paper lays the foundation for shared action, which includes many different stakeholders with differing aims. In this section, we outline key opportunities for future transdisciplinary engagement, identifying opportunities for scholars to continue building knowledge about dark patterns and their harms, for regulators and other enforcement agencies to better detect and thus sanction dark patterns, and for legal scholars and legislators to address current and future consequences of dark patterns that can inform further action.

\subsection{Challenges in Evolving the Ontology}
Not all of our mappings were clear-cut and some may be productively extended or disputed in future versions of this ontology. Through dialogue, we sought to locate existing patterns within our ontology based on our best understanding of the pattern as described by its name and definition in the source taxonomy. One challenge we faced was that some combinations of patterns have evolved over time. For instance, Mathur et al.'s~\cite{Mathur2019-hx} high-level pattern ``social proof'' originated with two sub-patterns, ``activity messages'' and ``testimonials.'' Later, the FTC created new low-level patterns, introducing ``endorsements'' (we bundled it with testimonials as one low-level pattern) and more specific types of endorsement or testimonials (e.g., ``deceptive celebrity endorsements,'' ``false activity messages''). Future work could identify the most useful level of abstraction for these patterns. Additionally, the use of novel names for patterns (particularly by the EDPB and CMA) or the use of patterns in specific contexts (e.g., e-commerce, social media) caused us to consider both the presence of granular low-level patterns and the relation of these low-level patterns to inferred meso-level patterns. In particular, the use of novel names for patterns types and definitions was a challenge from an analytic perspective, resulting in: 
i) instances where a wholly new pattern was introduced (e.g., CMA's ``information overload'' which we leveraged to create a new meso-level pattern of ``choice overload''); ii) instances where a new high-level strategy was highly similar to an existing high-level strategy (e.g., EDPB's ``skipping'' which we subsumed within ``sneaking''); and iii) instances where existing patterns included both a generalizable pattern and domain-specific information which may need to be captured in specific low-level patterns in future work (e.g., EDPB's ``left in the dark'' is a form of ``hidden information'' but implies specific low-level patterns that are specific to data protection). These observed challenges point towards the value of a shared ontology that includes a consistent vocabulary, but also points to opportunities to generate more specific knowledge that is linked to particular contexts and technologies. For instance, low-level patterns could be tagged based on how well they relate to specific contexts (e.g., e-commerce, social media), technologies (e.g., CUIs, VR/AR, robots), or application domains (e.g., health, travel) as indicated by a recent systematic review of dark patterns literature~\cite{Gray2023-systematic}. 

\subsection{Activating Transdisciplinary Pathways}
As we have outlined, work relating to dark patterns has connected many different disciplinary communities toward shared goals, including social scientists studying the presence and harms of dark patterns, legal scholars linking instances of dark patterns to relevant consumer protection or data protection legal frameworks, legislators targeting specific legal provisions about dark patterns to support new obligations and/or future sanctions, and regulators detecting legal violations related to dark patterns to support enforcement sanctions. We consider multiple opportunities for collaboration within and across these stakeholder groups:

\begin{itemize}
    \item \textbf{Social Scientists} Scientists studying dark patterns can use the ontology to better map the impact triggered by certain dark patterns in concrete contexts in ways that support shared knowledge building and reduce duplication. This approach has been applied for specific low-level patterns by various empirical studies that evaluated the impact of dark pattern design on the outcome of users' consent decisions~\cite{Biel-LINC-2023}, but could be scaled up substantially using the ontology as a means of producing and sharing these mappings. 
    
    \item \textbf{Social Scientists + Computer Scientists} The detection of dark patterns could also be more robustly supported by our ontology, with our assertion that low-level patterns show the most promise in being detectable. Existing detection efforts (e.g.,~\cite{Stavrakakis2021-hz,Soe2022-lt,Mathur2019-hx,Nouwens2020-ij,hofslot-2022-AutomaticClassifViolCookieBanners,chen2023-ut,Bouhoula-2023,Koch-2023,Yada2022-he}) have shown that higher-level patterns are difficult or impossible to detect at scale due to their abstract nature that requires interpretation, while low-level UI elements with discrete and known qualities (e.g., cookie consent banners, elements of the checkout process) are more detectable using software tools for automated detection. Our ontology of low-level patterns and gaps creates a foundation for future detection efforts, allowing computer science scholars to focus on pattern types which are most likely to be detectable and measurable.

    \item \textbf{Social Scientists + Regulators} Bielova et al.~\cite{Bielova2024-zr} have recently compared the results of such empirical studies and designs recommended by EU regulators and found multiple gaps and contradictions relating to instances of dark patterns, showing that empirical studies bring important insights not only in the research community but also for the regulators and policy-makers. This effort demonstrates an opportunity for regulators and social scientists to work more closely---commissioning studies where user experience of dark patterns is unknown or unclear (particularly with relation to causal mechanisms) while deprioritizing studies that address design choices that are already illegal under statute.

    \item \textbf{Social Scientists + Legal Scholars} The ontology can be extended to consider potential harms in relation to specific dark patterns types~\cite{Gunawan2022-harms}. For example, the meso-level dark pattern \textit{Nagging} can arguably trigger ``attentional theft,'' thus harming consumer welfare, and can lead to indirect harms such as increased vulnerability to privacy violations, and finally, to anti-competitive harms~\cite{Hung-nagging}. A mapping of harms to specific types of dark patterns in the ontology may support connections to avenues for legal remedies, as well as aid in identifying areas where additional research is needed.

    \item \textbf{Legal Scholars + Regulators} The ontology may also be extended to refer to concrete enforcement cases already consolidated in a database of dark patterns case law, such as those on Brignull's updated site~\cite{deceptivedesign-legalcases}. This will allow for case law to inform future legal sanctions, identify which elements of the ontology connect to existing legal frameworks, and lay the groundwork for future legislative action to allow for sanctioning of novel patterns that are not well addressed through existing laws.

\end{itemize}

\section{Conclusion}
To support the development of a shared language of dark patterns, 
in this paper we present our analysis of ten existing regulatory and academic taxonomies of dark patterns and propose a three-level ontology with standardized definitions for 65 synthesized dark pattern types across low-, meso-, and high-level patterns. Building on our analysis, future scholars, regulators, and legal professionals can benefit from our hierarchical organization of dark patterns types to indicate links to existing and similar concepts. This description encourages the establishment of provenance in future work, allowing scholars and regulators to identify pattern types and their origins and provide an audit trail to connect specific contextually-bound instances with broader categorizations. This ontology creates a foundation for a shared and reusable knowledge source, allowing many stakeholders to work together in building a shared, explicit and precise conceptualization of what is already known in the literature and which can be further refined and extended. Finally, we illustrate how this ontology can support translational research and regulatory action,  by extending the ontology from three contemporary studies defining dark patterns from domain and context-specific areas, as well as ontology extension to map legislation and case law.

\begin{acks}
This work is funded in part by the National Science Foundation under Grant No. 1909714 and the ANR 22-PECY-0002 IPOP (Interdisciplinary Project on Privacy) project of the Cybersecurity PEPR.
\end{acks}
\balance
\bibliographystyle{ACM-Reference-Format}
\bibliography{evolution}

\newpage
\appendix

\section{Final Ontology Definitions}

\begin{itemize}

\item
  \textbf{Sneaking} is a strategy which hides, disguises, or delays the
  disclosure of important information that, if made available to users,
  would cause a user to unintentionally take an action they would likely
  object to.

  \begin{itemize}
  \item
    \textbf{Bait and Switch} subverts the user's expectation that their
    choice will result in a desired action, instead leading to an
    unexpected, undesirable outcome.

    \begin{itemize}
    \item
      \textbf{Disguised Ads} \emph{Bait and Switch} and use
      \emph{Sneaking} to style interface elements so they are not
      clearly marked as an advertisement or other biased source. As a
      result, users are induced into clicking on the interface element
      because they assume that it is a relevant and salient interaction,
      leading to unwitting interaction with advertising content.
    \end{itemize}
  \item
    \textbf{Hiding Information} subverts the user's expectation that all
    relevant information to make an informed choice will be available to
    them, instead hiding information or delaying the disclosure of
    information until later in the user journey that may have led to
    them making another choice.

    \begin{itemize}
    \item
      \textbf{Sneak into Basket} \emph{Hides Information} and uses
      \emph{Sneaking} to add unwanted items to a user's shopping cart
      without their consent. As a result, a user assumes that only the
      items they explicitly added to their cart will be purchased,
      leading to unintentional purchase of additional items.
    \item
      \textbf{Drip Pricing, Hidden Costs, or Partitioned Pricing}
      \emph{Hides Information} and uses \emph{Sneaking} to reveal new
      charges or costs, present only partial price components, or
      otherwise delay revealing the full price of a product or service
      through late or incomplete disclosure. As a result, the user is
      misled about the total or complete price of the product or
      service, leading to them to make a purchase decision after they
      have expended effort on false pretenses.
    \item
      \textbf{Reference Pricing} \emph{Hides Information} and uses
      \emph{Sneaking} to include a misleading or inaccurate price for a
      product or service that makes a discounted price appear more
      attractive. As a result, the user is misled into believing that
      the price they pay is discounted, leading them to make a decision
      to purchase a product or service on false pretenses.
    \end{itemize}
  \item
    \textbf{(De)contextualizing Cues} subverts the user's expectation
    that provided information will guide the user to making an informed
    choice, instead confusing the user and/or preventing them from
    locating relevant information due to the context where information
    is presented.

    \begin{itemize}
    \item
      \textbf{Conflicting Information} uses \emph{(De)contextualizing
      Cues} and \emph{Sneaking} to include two or more sources of
      information that conflict with each other. As a result, the user
      is unsure what the consequences of their actions will be and will
      be more likely to accept default settings that may not be in their
      best interest.
    \item
      \textbf{Information without context} uses
      \emph{(De)contextualizing Cues} and \emph{Sneaking} to alter the
      relevant information or user controls to limit discoverability. As
      a result, the user is unlikely to find the information or action
      possibility they are interested in.
    \end{itemize}
  \end{itemize}
\item
  \textbf{Obstruction} is a strategy which impedes a user's task flow,
  making an interaction more difficult than it inherently needs to be,
  dissuading a user from taking an action.

  \begin{itemize}
  \item
    \textbf{Roach Motel} subverts the user's expectation that an action
    will be as easy to reverse as it is to make, instead creating a
    situation that is easy to get into, but difficult to get out of.

    \begin{itemize}
    \item
      \textbf{Immortal Accounts} create a \emph{Roach Motel} and use
      \emph{Obstruction} to make it difficult or impossible to delete a
      user account once it has been created. As a result, the user may
      create an account or share data with the false assumption that
      they can later delete this information, even though that account
      and/or data are then unable to be removed by the user.
    \item
      \textbf{Dead Ends} create a \emph{Roach Motel} and use
      \emph{Obstruction} to prevent users from finding information
      through inactive links or redirections that limit or completely
      prevent the display of relevant information. As a result, the user
      may seek to find relevant information or action possibilities but
      instead be left unable to achieve their goal.
    \end{itemize}
  \item
    \textbf{Creating Barriers} subverts the user's expectation that
    relevant user tasks will be supported by the interface, instead
    preventing, abstracting, or otherwise complicating a user task to
    disincentive user action.

    \begin{itemize}
    \item
      \textbf{Price Comparison Prevention} \emph{Creates Barriers} and
      uses \emph{Obstruction} by excluding relevant information,
      limiting the ability of a user to copy/paste, or otherwise
      inhibiting a user from comparing prices across two or more
      vendors. As a result, the user cannot make an informed decision
      about where to buy a product or service.
    \item
      \textbf{Intermediate Currencies} \emph{Create Barriers} and use
      \emph{Obstruction} to hide the true cost of a product or service
      by requiring the user to spend real money to purchase a virtual
      currency that is then used to purchase a product or service. As a
      result, the user is unable to easily ascertain the true monetary
      cost of a product or service, leading them to make an uninformed
      purchase decision based on an obscured cost.
    \end{itemize}
  \item
    \textbf{Adding Steps} subverts the user's expectation that a task
    will take as few steps as technologically needed, instead creating
    additional points of unnecessary but required user interaction to
    perform a task.

    \begin{itemize}
    \item
      \textbf{Privacy Mazes} \emph{Add Steps} and use \emph{Obstruction}
      to require a user to navigate through many pages to obtain
      relevant information or control without a comprehensive and
      exhaustive overview. As a result, the user is prevented from
      easily discovering relevant information or action possibilities,
      leaving them unable to make informed decisions regarding their
      privacy.
    \end{itemize}
  \end{itemize}
\item
  \textbf{Interface Interference} is a strategy which privileges
  specific actions over others through manipulation of the user
  interface, thereby confusing the user or limiting discoverability of
  relevant action possibilities.

  \begin{itemize}
  \item
    \textbf{Manipulating Visual Choice Architecture} subverts the user's
    expectation that the options presented will support their desired
    goal, instead including an order or structure of options that makes
    another outcome more likely.

    \begin{itemize}
    \item
      \textbf{False Hierarchy} \emph{Manipulates the Visual Choice
      Architecture,} using \emph{Interface Interference} to give one or
      more options visual or interactive prominence over others,
      particularly where items should be in parallel rather than
      hierarchical. As a result, the user may misunderstand or be unable
      to accurately compare their options, making a selection based on a
      false or incomplete choice architecture.
    \item
      \textbf{Visual Prominence} \emph{Manipulates the Visual Choice
      Architecture,} using \emph{Interface Interference} to place an
      element relevant to user goals in visual competition with a more
      distracting and prominent element. As a result, the user may
      forget about or be distracted from their original goal, even if
      that goal was their primary intent.
    \item
      \textbf{Bundling} \emph{Manipulates the Visual Choice
      Architecture,} using \emph{Interface Interference} to group two or
      more products or services in a single package at a special price.
      As a result, the user may incorrectly assume that these items must
      be purchased as a bundle or be unaware of the unbundled price for
      the component elements, possibly leading to an uninformed
      purchasing decision.
    \item
      \textbf{Pressured Selling} \emph{Manipulates the Visual Choice
      Architecture,} using \emph{Interface Interference} to preselect or
      use visual prominence to focus user attention on more expensive
      product options . As a result, the user may be unaware that a
      lower price is available or even desirable for their needs ,
      steering the user into making a more expensive product selection
      than they otherwise would have.
    \end{itemize}
  \item
    \textbf{Bad Defaults} subverts the user's expectation that default
    settings will be in their best interest, instead requiring users to
    take active steps to change settings that may cause harm or
    unintentional disclosure of information.
  \item
    \textbf{Emotional or Sensory Manipulation} subverts the user's
    expectation that the design of the site will allow them to achieve
    their goal without manipulation, instead altering the language,
    style, color, or other design elements to evoke an emotion or
    manipulate the senses in order to persuade the user into a
    particular action.

    \begin{itemize}
    \item
      \textbf{Cuteness} uses \emph{Emotional or Sensory Manipulation}
      and \emph{Interface Interference} to embed attractive cues in the
      design of a robot interface or form factor. As a result, a user
      may place undue trust in the robot, leading the user to
      inaccurately or incompletely assess the risks of interacting with
      the robot.
    \item
      \textbf{Positive or Negative Framing} uses \emph{Emotional or
      Sensory Manipulation} and \emph{Interface Interference} to
      visually obscure, distract, or persuade a user from important
      information they need to achieve their goal. As a result, the user
      may assume that the system is providing equal access to relevant
      information, leading the user to be distracted by positive or
      negative aesthetic cues that distract them from important
      information or action possibilities or otherwise convince them to
      pursue a different goal.
    \end{itemize}
  \item
    \textbf{Trick Questions} subvert the user's expectation that prompts
    will be written in a straightforward and intelligible manner,
    instead using confusing wording, double negatives, or otherwise
    leading language or interface cues to manipulate a user's choice.
  \item
    \textbf{Choice Overload} subverts the user's expectation that the
    choices they make should be understandable and comparable, instead
    providing too many options to compare or encouraging users to
    overlook relevant information due to the volume of choices provided.
  \item
    \textbf{Hidden Information} subverts the user's expectation that
    relevant information will be made accessible and visible, instead
    disguising relevant information or framing it as irrelevant.
  \item
    \textbf{Language Inaccessibility} subverts the user's expectation
    that guidance will be provided in a way that is understandable and
    intelligible, instead using unnecessarily complex language or a
    language not spoken by the user to decrease the likelihood the user
    will make an informed choice.

    \begin{itemize}
    \item
      \textbf{Wrong Language} leverages \emph{Language Accessibility},
      using \emph{Interface Interference} to provide important
      information in a different language than the official language of
      the country where users live. As a result, the user will not have
      access to relevant information about their interaction with the
      system and their ability to choose, leading to uninformed
      decisions.
    \item
      \textbf{Complex Language} leverages \emph{Language Accessibility},
      using \emph{Interface Interference} to make information difficult
      to understand by using obscure word choices and/or sentence
      structure. As a result, the user will not be able to comprehend
      relevant information about their interaction with the system and
      their ability to choose, leading to uninformed decisions.
    \end{itemize}
  \item
    \textbf{Feedforward Ambiguity} subverts the user's expectation that
    their choice will be likely to result in an action they can predict,
    instead providing a discrepancy between information and actions
    available to users that results in an outcome that is different from
    what the user expects.
  \end{itemize}
\item
  \textbf{Forced Action} is a strategy which requires users to perform
  an additional and/or tangential action or information to access (or
  continue to access) specific functionality, preventing them from
  continuing their interaction with a system without performing that
  action.

  \begin{itemize}
  \item
    \textbf{Nagging} subverts the user's expectation that they have rational control over the interaction they make with a system, instead distracting the user from a desired task the user is focusing on to induce an action or make a decision the user does not want to make by repeatedly interrupting the user during normal interaction.
    \item \textbf{Forced Continuity} subverts the user's expectation that a
    subscription created in the past will not auto-renew or otherwise
    continue in the future, instead causing undesired charges,
    difficulty to cancel, or lack of awareness that a subscription is
    still active.
  \item
    \textbf{Forced Registration} subverts the user's expectation that
    they can complete an action without registering or creating an
    account, instead tricking them into thinking that registration is
    required, often resulting in the sharing of unneeded personal data.
  \item
    \textbf{Forced Communication or Disclosure} subverts the user's
    expectation that a system will only request information needed to
    complete their desired goals, instead tricking them into sharing
    more information about themselves or using their information for
    purposes that they do not desire.

    \begin{itemize}
    \item
      \textbf{Privacy Zuckering} uses \emph{Forced Communication or
      Disclosure} as a type of \emph{Forced Action} to trick users into
      sharing more information about themselves than they intend to or
      would agree to if fully informed. As a result, the user assumes
      that information they are requested to provide is vital for use of
      the service, even while this information is used or sold for other
      purposes.
    \item
      \textbf{Friend Spam} uses \emph{Forced Communication or
      Disclosure} as a type of \emph{Forced Action} to collect
      information about other users through extractive means that
      results in unwanted contact from the service. As a result, the
      user assumes that information about their friends or social
      network is vital for use of the service, even while this
      information is used to spam other users.
    \item
      \textbf{Address Book Leeching} uses \emph{Forced Communication or
      Disclosure} as a type of \emph{Forced Action} to collect
      information about other users through extractive means, which are
      often hidden to the user and/or conducted under false pretenses.
      As a result, the user assumes that only vital information will be
      collected when signing up for or using a service, even while this
      information is used to gain knowledge of other users or inform
      other purposes that have not been initially declared.
    \item
      \textbf{Social Pyramid} uses \emph{Forced Communication or
      Disclosure} as a type of \emph{Forced Action} to manipulate
      existing users into recruiting new users to use a service, often
      by tying this recruitment to additional functionality or other
      benefits. As a result, the user assumes that social recruiting is
      necessary to continue to use aspects of the service, even while
      this information is primarily used to build the service's user
      base.
    \end{itemize}
  \item
    \textbf{Gamification} subverts the user's expectation that system
    functionality is based on alignment with user goals and needs,
    instead coercing them into gaining access to aspects of a service
    through repeated (and perhaps undesired) use of aspects of the
    service.

    \begin{itemize}
    \item
      \textbf{Pay-to-Play} uses \emph{Gamification} as a type of
      \emph{Forced Action} to initially claim that aspects of a service
      or product are available via purchase or download, but then later
      charging users to actually obtain that functionality. As a result,
      the user incorrectly assumes that a service or product will allow
      them certain functionality, leading to them downloading or
      purchasing the product or service under false pretenses.
    \item
      \textbf{Grinding} uses \emph{Gamification} as a type of
      \emph{Forced Action} to require repeated, often cumbersome and
      labor-intensive actions over time in order to obtain certain
      relevant functionality. As a result, the user may seek to avoid
      these repetitive actions, leading to them making unwanted
      additional in-app purchases to unlock the same functionality
      without ``grinding'' over an extended period of time.
    \end{itemize}
  \item
    \textbf{Attention Capture} subverts the user's expectation that they
    have rational control over the time they spend using a system,
    instead tricking them into spending more time or other resources to
    continue use for longer than they otherwise would.

    \begin{itemize}
    \item
      \textbf{Auto-Play} uses \emph{Attention Capture} as a type of
      \emph{Forced Action} to automatically play new video after an
      existing video has completed. As a result, the user may lose
      control over their viewing experience, leading them to watch more
      content than they intended or result in them watching content that
      is unexpected or harmful.
    \end{itemize}
  \end{itemize}
\item
  \textbf{Social Engineering} is a strategy which presents options or
  information that causes a user to be more likely to perform a specific
  action based on their individual and/or social cognitive biases,
  thereby leveraging a user's desire to follow expected or imposed
  social norms.

  \begin{itemize}
  \item
    \textbf{Scarcity or Popularity Claims} subverts the user's
    expectation that information provided about a product's availability
    or desirability is accurate , instead pressuring the user to
    purchase a product without additional reflection or verification.

    \begin{itemize}
    \item
      \textbf{High Demand} uses \emph{Scarcity and Popularity Claim}s as
      a type of \emph{Social Engineering} to indicate that a product is
      in high-demand or likely to sell out soon, even though that claim
      is misleading or false. As a result, the user may assume that
      demand is high when it is not, leading to their uninformed
      purchase of a product or service.
    \end{itemize}
  \item
    \textbf{Social Proof} subverts the user's expectation that the
    indicated behavior of others in a specific situation is correct or
    desirable, instead accelerating user decision-making and encouraging
    the user to trust flawed implications through provided information.

    \begin{itemize}
    \item
      \textbf{Low Stock} uses \emph{Social Proof} as a type of
      \emph{Social Engineering} to indicate that a product is limited in
      quantity, even though that claim is misleading or false. As a
      result, the user may assume that a product is desirable due to
      demand, leading to undue or uninformed pressure to buy the product
      immediately.
    \item
      \textbf{Endorsements and Testimonials} use \emph{Social Proof} as
      a type of \emph{Social Engineering} to indicate that a product or
      service has been endorsed by another consumer, even though the
      source of that endorsement or testimonial is biased, misleading,
      incomplete, or false. As a result, the user may assume that the
      endorsement or testimonial is accurate and unbiased, leading to
      their uninformed purchase of a product or service.
    \item
      \textbf{Parasocial Pressure} uses \emph{Social Proof} as a type of
      \emph{Social Engineering} to indicate that a product or service
      has been endorsed by a celebrity, influencer, or other entity that
      the user trusts, even though the source of that endorsement is
      biased, misleading, incomplete, or false. As a result, the user
      may assume that the endorsement is accurate and unbiased, leading
      to their uninformed purchase of a product or service.
    \end{itemize}
  \item
    \textbf{Urgency} subverts the user's expectation that information
    provided about discounts or a limited-time deal for a product is
    accurate, instead accelerating the user's decision-making process by
    demanding immediate or timely action.

    \begin{itemize}
    \item
      \textbf{Activity Messages} use \emph{Urgency} as a type of
      \emph{Social Engineering} to describe other user activity on the
      site or service, even though the data presented about other users'
      purchases, views, visits, or contributions are misleading or
      false. As a result, the user may falsely feel a sense of urgency,
      assuming that others users are purchasing or otherwise interested
      product or service, leading to their uninformed purchase of a
      product or service.
    \item
      \textbf{Countdown Timers} use \emph{Urgency} as a type of
      \emph{Social Engineering} to indicate that a deal or discount will
      expire by displaying a countdown clock or timer, even though the
      clock or timer is completely fake, disappears, or resets
      automatically. As a result, the user may feel undue urgency and
      purchasing pressure, leading to their uninformed purchase of a
      product or service.
    \item
      \textbf{Limited Time Messages} use \emph{Urgency} as a type of
      \emph{Social Engineering} to indicate that a deal or discount will
      expire soon or be available only for a limited time, but without
      specifying a specific deadline. As a result, the user may feel
      undue urgency and purchasing pressure, leading to their uninformed
      purchase of a product or service.
    \end{itemize}
  \item
    \textbf{Personalization} subverts the user's expectation that
    products or service features are offered to all users in similar
    ways, instead using personal data to shape elements of the user
    experience that manipulate the user's goals while hiding other
    alternatives.

    \begin{itemize}
    \item
      \textbf{Confirmshaming} uses \emph{Personalization} as a type of
      \emph{Social Engineering} to frame a choice to opt-in or opt-out
      of a decision through emotional language or imagery that relies
      upon shame or guilt. As a result, the user may be convinced to
      change their goal due to the emotionally manipulative tactics,
      resulting in being steered away from making a choice that matched
      their initial goal.
    \end{itemize}
  \end{itemize}
\end{itemize}

\section{Analyzed Taxonomies of Dark Patterns}
\begin{table*}[]
\caption{Academic taxonomies of dark patterns.}
\label{tab:academic}
\begin{tabularx}{\textwidth}{llX}
\toprule
                               & High-Level Pattern             & Low-Level Pattern                                                                                                                                                                                           \\
\midrule
Brignull 2018-2022~\cite{Brignull_undated-vj}                       & —                              & Sneak into Basket, Bait and Switch, Roach Motel, Price Comparison Prevention, Disguised Ads, Privacy Zuckering, Trick Questions, Hidden Costs, Confirmshaming, Friend Spam, Forced Continuity, Misdirection \\
\midrule
Brignull 2023~\cite{Brignull_2023}                       & —                              & Comparison Prevention, Confirmshaming, Disguised Ads, Fake Scarcity, Fake Social Proof, Fake Urgency, Forced Action, Hard to Cancel, Hidden Costs, Hidden Subscription, Nagging, Obstruction, Preselection, Sneaking, Trick Wording, Visual Interference \\
\midrule
\multirow{6}{*}{Bösch et al.~\cite{Bosch2016-vc}}  & Obscure                        & Privacy Zuckering, Immortal Accounts, Hidden Legalese Stipulations, Bad Defaults                                                                                                                            \\
                               & Maximize                       & Shadow User Profiles,  Address Book Leeching, Forced Registration                                                                                                                                           \\
                               & Deny                           & Immortal Accounts                                                                                                                                                                                           \\
                               & Preserve                       & Shadow User Profiles,  Address Book Leeching                                                                                                                                                                \\
                               & Centralize                     & Shadow User Profiles                                                                                                                                                                                        \\
                               & Publish, Violate, Fake         & —                                                                                                                                                                                                           \\
\midrule
\multirow{5}{*}{Gray et al.~\cite{Gray2018-or}}   & Nagging                        & —                                                                                                                                                                                                           \\
                               & Sneaking                       & Intermediate-Level Currency, Roach Motel, Price Comparison Prevention                                                                                                                                       \\
                               & Obstruction                    & Bait and Switch, Sneak into Basket, Hidden Costs, Forced Continuity                                                                                                                                         \\
                               & Interface Interference         & Toying with Emotion, Aesthetic Manipulation, Trick Questions, Preselection, Disguised Ad, Hidden Information, False Hierarchy                                                                               \\
                               & Forced Action                  & Gamification, Privacy Zuckering, Social Pyramid                                                                                                                                                             \\
\midrule
\multirow{7}{*}{Mathur et al.~\cite{Mathur2019-hx}} & Sneaking                       & Sneak into Basket, Hidden Costs, Hidden Subscription                                                                                                                                                        \\
                               & Urgency                        & Limited-time Message, Countdown Timer                                                                                                                                                                       \\
                               & Misdirection                   & Confirmshaming, Visual Interference, Trick Questions, Pressured Selling                                                                                                                                     \\
                               & Social Proof                   & Activity Message, Testimonials                                                                                                                                                                              \\
                               & Scarcity                       & Low-stock Message, High-demand Message                                                                                                                                                                      \\
                               & Obstruction                    & Hard to Cancel                                                                                                                                                                                              \\
                               & Forced Action                  & Forced Enrollment                                                                                                                                                                                           \\
\midrule
\multirow{8}{*}{Luguri et al.~\cite{Luguri2021-bg}} & Nagging                        & —                                                                                                                                                                                                           \\
                               & Social Proof                   & Testimonials, Activity Messages                                                                                                                                                                             \\
                               & Obstruction                    & Immortal Accounts, Intermediate-Level Currency, Roach Motel, Price Comparison Prevention                                                                                                                    \\
                               & Sneaking                       & Bait and Switch, Sneak into Basket, Hidden Costs, Hidden Subscription / Forced Continuity                                                                                                                   \\
                               & Interface Interference         & Cuteness, False Hierarchy / Pressured Selling, Toying with Emotion, Trick Questions, Preselection, Disguised Ad, Hidden Information / Aesthetic Manipulation, Confirmshaming                                \\
                               & Forced Action                  & Friend spam/social pyramid/address book leeching, Privacy Zuckering, Gamification, Forced Registration                                                                                                      \\
                               & Scarcity                       & High Demand Message, Low Stock Message                                                                                                                                                                      \\
                               & Urgency                        & Countdown Timer, Limited Time Message                                                                                                                                                                       \\
\bottomrule                               
\end{tabularx}
\end{table*}

\begin{table*}[]
\caption{Regulatory taxonomies of dark patterns. }
\label{tab:regulatory}
\begin{tabularx}{\textwidth}{p{2.2cm}p{2.3cm}X}
\toprule
                               & High-Level \hspace{0.5cm} Pattern             & Low-Level Pattern                                                                                                                                                                                           \\
\midrule
\multirow{6}{*}{EDPB~\cite{EDPB2023}}          & Overloading                    & Continuous Prompting, Privacy Maze, Too Many Options                                                                                                                                                        \\
                               & Skipping                       & Deceptive Snugness, Look Over There                                                                                                                                                                         \\
                               & Stirring                       & Emotional Steering, Hidden in Plain Sight                                                                                                                                                                   \\
                               & Obstructing                      & Dead End, Longer than Necessary, Misleading Action                                                                                                                                                     \\
                               & Fickle                         & Lacking Hierarchy, Decontextualizing, Language Discontinuity, Inconsistent Interface                                                                                                                                                                        \\
                               & Left in the Dark               & Conflicting Information, Ambiguous Wording or Information                                                                                                                           \\
\midrule
\multirow{7}{*}{EU Com. (EC)~\cite{EUCOM2022}} & Nagging                        & —                                                                                                                                                                                                           \\
                               & Social Proof                   & Testimonials, Activity Messages                                                                                                                                                                             \\
                               & Obstruction                    & Intermediate-Level Currency, Roach Motel / Difficult Cancellations, Price Comparison Prevention                                                                                                             \\
                               & Sneaking                       & Bait and Switch, Sneak into Basket, Hidden Costs, Hidden Subscription / Forced Continuity                                                                                                                   \\
                               & Interface Interference         & Toying with Emotion, Trick Questions, Preselection (default), Disguised Ad, Hidden Information / False Hierarchy, Confirmshaming                                                                            \\
                               & Forced Action                  & Forced Registration                                                                                                                                                                                         \\
                               & Urgency                        & Countdown Timer / Limited TIme Message, Low Stock / High Demand Message                                                                                                                                     \\
\midrule
\multirow{8}{*}{OECD~\cite{Oecd2022}}        & Forced Action    & Forced Registration, Forced Disclosure / Privacy Zuckering, Friend Spam / Social Pyramid / Address Book Leeching, Gamification                                                                                \\
                               & Interface Interference                       & Hidden Information, False Hierarchy, Preselection, Misleading Reference Pricing, Trick Questions, Disguised Ads, Confirmshaming / Toying with Emotion                                                                                                                                                          \\
                               & Nagging                        & Nagging                                                                                                                                 \\
                               & Obstruction                    & Hard to Cancel or Opt Out / Roach Motel / Click Fatigue / Ease, (Price) Comparison Prevention, Immortal Accounts, Intermediate Currency                                                                                                                                    \\
                               & Sneaking & Sneak into Basket, Hidden Costs / Drip Pricing, Hidden Subscription / Forced Continuity, Bait and Switch (including Bait Pricing)                                                                           \\
                               & Social Proof                & Activity Messages, Testimonials                                   \\
                               & Urgency              & Low Stock / High Demand Message, Countdown Timer / Limited Time Message  \\ 
\midrule
\multirow{3}{*}{UK CMA~\cite{CMA2022}}        & Choice Structure               & Defaults, Ranking, Partitioned Pricing, Sludge, Bundling, Dark nudge, Choice overload and decoys, Virtual currencies in gaming, Sensory manipulation, Forced outcomes                                       \\
                               & Choice Information             & Drip pricing, Reference pricing, Framing, Complex language, Information overload                                                                                                                            \\
                               & Choice Pressure                & Scarcity and popularity claims, Prompts and reminders, Messengers, Commitment, Feedback, Personalisation                                                                                                    \\
\midrule
\multirow{8}{*}{US FTC~\cite{FTC2022}}        & Endorsements (Social Proof)    & False Activity Messages, Deceptive Consumer Testimonials, Deceptive Celebrity Endorsements, Parasocial Relationship Pressure                                                                                \\
                               & Scarcity                       & False Low Stock Message, False High Demand Message                                                                                                                                                          \\
                               & Urgency                        & False Discount Claims, False Limited Time Message, Baseless Countdown Timer                                                                                                                                 \\
                               & Obstruction                    & Immortal Accounts Roadblocks to Cancellation, Price Comparison Prevention                                                                                                                                   \\
                               & Sneaking or Information Hiding & Intermediate Currency, Hidden Subscription or Forced Continuity, Drip Pricing, Hidden Costs, Hidden Information, Sneak-into-Basket                                                                          \\
                               & Interface Interference         & Bait and Switch, Disguised Ads, False Hierarchy or Pressured Upselling, Misdirection                                                                                                                        \\
                               & Coerced Action                 & Friend Spam, Social Pyramid Schemes, and Address Book Leeching, Pay-to-Play or Grinding, Forced Registration or Enrollment, Nagging, Auto-Play, Unauthorized Transactions                                   \\
                               & Asymmetric Choice              & Subverting Privacy Preferences, Preselection, Confirm Shaming, Trick Questions  \\

\bottomrule                               
\end{tabularx}
\end{table*}

\end{document}